\documentclass[aps,prb,twocolumn,superscriptaddress,eqsecnum,showpacs,amsmath,amssymb, 10pt]{revtex4-1}
\usepackage{graphicx}
\usepackage{natbib}
\usepackage{color}
\definecolor{dblue}{rgb}{0.24,0.19,0.60}
\definecolor{turquoise}{rgb}{0.0,0.7,0.7}
\definecolor{lavande}{rgb}{0.21,0.2,0.59}
\definecolor{violine}{rgb}{0.63,0.3,0.07}
\definecolor{dgreen}{rgb}{0.05,0.70,0.05}

\definecolor{grayJ3}{rgb}{0.7,0.7,0.7}
\definecolor{grayJ2}{rgb}{0.4,0.4,0.4}

\newcommand{\beqa}{\begin{eqnarray}}
\newcommand{\eneqa}{\end{eqnarray}}
\newcommand{\beq}{\begin{equation}}
\newcommand{\eneq}{\end{equation}}
\newcommand{\nn}{\nonumber}
\newcommand{\ua}{\uparrow}
\newcommand{\da}{\downarrow}

\begin{document}

\title{Ferromagnetic frustrated spin systems on the square lattice: a Schwinger boson study}

\author{H.\ Feldner}
\affiliation{Institut de Physique et Chimie des Mat\'eriaux de Strasbourg, UMR7504, CNRS-UdS,
23 rue du Loess, BP43, 67034 Strasbourg Cedex 2, France}
\affiliation{Institut f\"ur Theoretische Physik, Georg-August-Universit\"at
G\"ottingen, Friedrich-Hund-Platz 1, 37077 G\"ottingen, Germany}

\author{D.C.\ Cabra}
\affiliation{Institut de Physique et Chimie des Mat\'eriaux de Strasbourg, UMR7504, CNRS-UdS,
23 rue du Loess, BP43, 67034 Strasbourg Cedex 2, France}
\affiliation{IFLP - Departamento de F\'{\i}sica, Universidad Nacional de La Plata, C.C. 67, 1900 La Plata, Argentina}

\author{G.L.\ Rossini}
\affiliation{IFLP - Departamento de F\'{\i}sica, Universidad Nacional de La Plata, C.C. 67, 1900 La Plata, Argentina}


\begin{abstract}

We study a ferromagnetic Heisenberg spin system on the square lattice, with nearest neighbors interaction $J_1$ frustrated by second $J_2$ and
third $J_3$ neighbors antiferromagnetic interactions, using a mean field theory for the Schwinger boson
representation of spins. For $J_3=0$ we find that the boundary between the ferromagnetic and the collinear
classical phases shifts to smaller values of $J_2$ when quantum fluctuations are included.
Along the line $J_2/\vert J_1\vert = 1$ the boundaries between the collinear and incommensurate regions
are strongly shifted to larger values with respect to the classical case.
We do not find clear evidence for spin gapped phases within the present approximation.

\end{abstract}

\pacs{
75.10.Jm; 
75.30.Kz; 
75.30.Ds. 
}
\maketitle


\section{Introduction}

Most current works on frustrated magnetic systems generally deal with competing antiferromagnetic interactions.
Recently, some frustrated systems have been discovered where the basic interaction is ferromagnetic.
In particular some vanadate and cuprate crystals as $(CuCl)LaNb_2O_7$,\cite{exp6} $Pb_2VO(PO_4)_2$,\cite{exp1,exp2,exp3,exp4,exp5}
 $SrZnVO(PO_4)_2$, \cite{exp5,exp7,exp8} $BaCdVO(PO_4)_2, $\cite{exp2,exp7,exp9}
and $PbZnVO(PO_4)_2$\cite{exp10} can be described by a two dimensional Heisenberg model of spin $S=1/2$
with a ferromagnetic first neighbors interaction and antiferromagnetic further neighbors interactions.
Other possible relevant materials are $ (CuBr)LaNb_2 O_7 $\cite{Uemura} which shows collinear order, 
and $(CuBr)Sr_2 Nb_3 O_{10} $\cite{Tsujimoto} which shows a plateau at $M=1/3$ in the magnetization curve. 
As one dimensional counterparts, materials like $LiCuVO_4$ \cite{enderle05,enderle10} and $Li_2ZrCuO_4$ \cite{drechsler07} can be modeled by
ferromagnetic frustrated spin $S=1/2$  Heisenberg chains.

\begin{figure}[h!t]
\centering
\includegraphics[width=1\columnwidth]{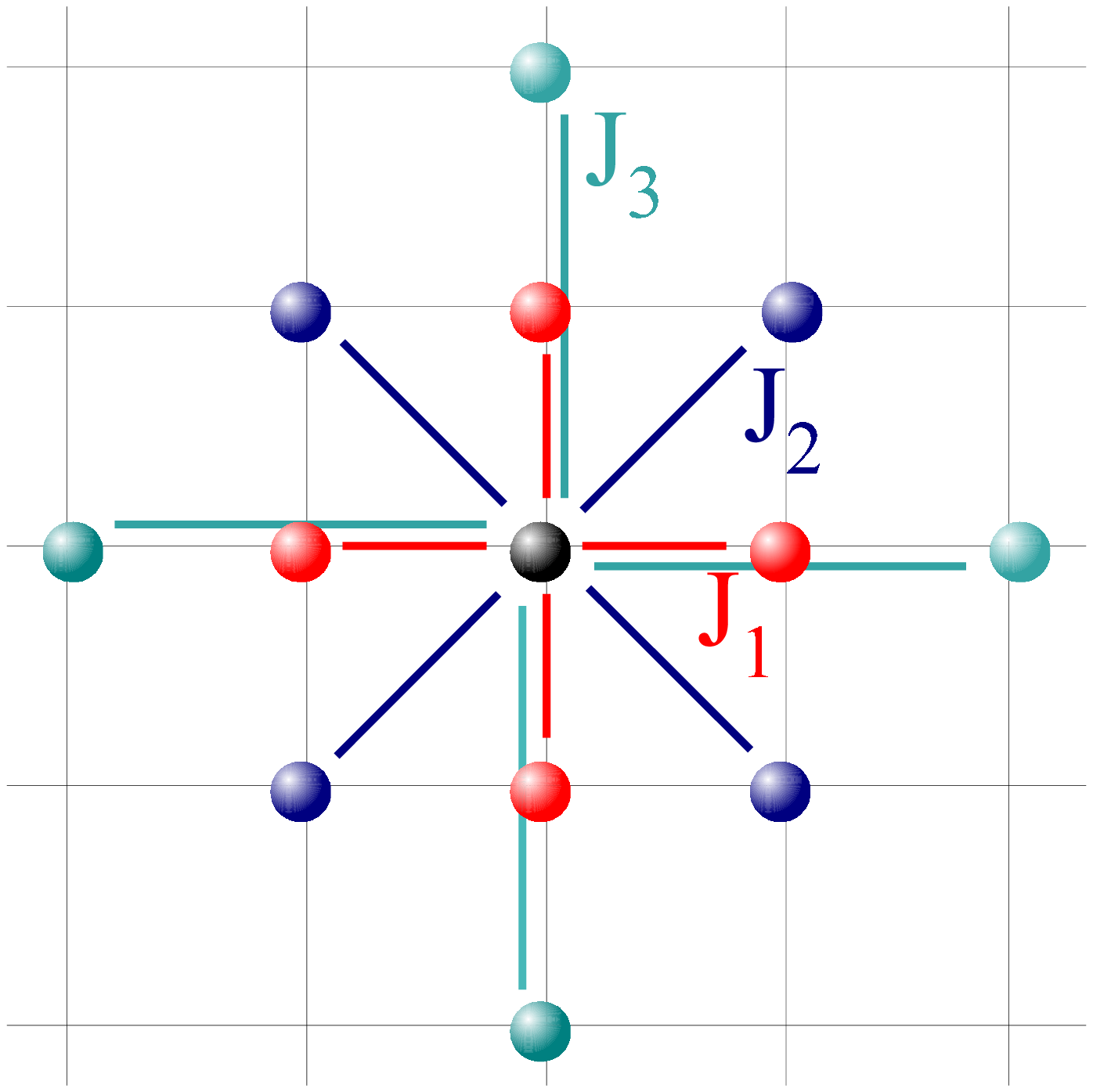}
\caption{(color online) Square lattice with $J_1<0$ ferromagnetic
first neighbors couplings, $J_2>0$ antiferromagnetic second neighbors
couplings and $J_3>0$ antiferromagnetic third neighbors couplings.}
 \label{fig:model}
\end{figure}

In the present work we consider such a two dimensional Heisenberg model on the square lattice (see  Fig.\ \ref{fig:model}) with
ferromagnetic nearest neighbors 
interactions $J_1<0$, frustrated by next to nearest neighbors 
antiferromagnetic interactions $ J_2>0$ and also 
third neighbors antiferromagnetic interactions $ J_3>0$, given by the Hamiltonian
\beqa
\label{eq:H}
H & = & J_1\sum_{\langle i,j\rangle_1} \vec{S}_i. \vec{S}_j
+ J_2\sum_{ \langle i,j\rangle_2}  \vec{S}_i \cdot \vec{S}_j\nn\\
&& + J_3\sum_{\langle i,j\rangle_3} \vec{S}_i\cdot  \vec{S}_j
\eneqa
where $\vec{S}_i$ is the spin $S$ operator at site $i$ and the range of the interacting neighbor
sites $i,j$ is indicated by brackets $\langle i,j\rangle_r$
with $r=1,2,3$. Closely related work has been done for antiferromagnetic $J_1$.\cite{Jolicoeur,Ferrer,Arlego}

\begin{figure}[h!t]
\centering
\includegraphics[width=1\columnwidth]{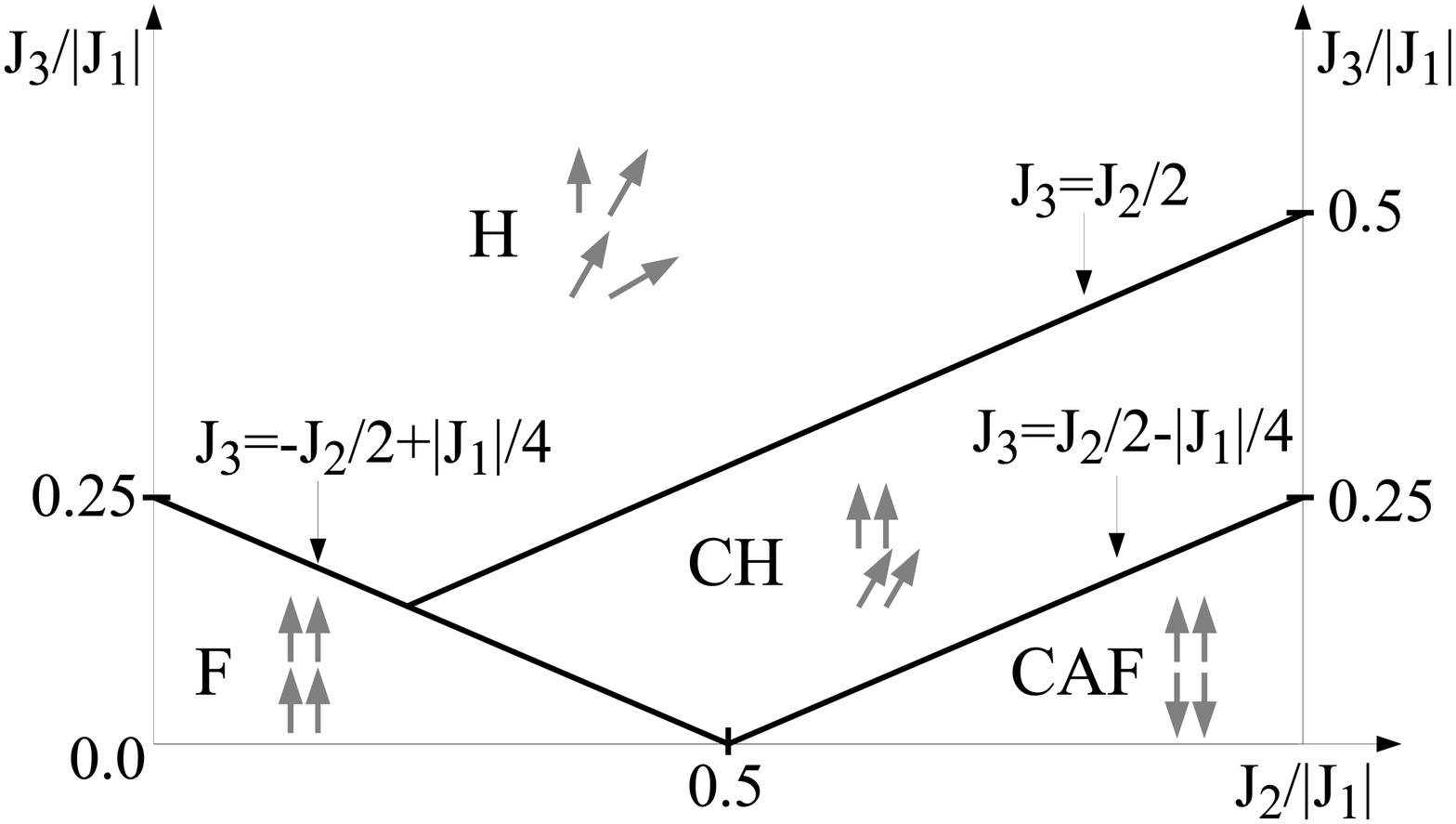}
\caption{
Classical phase diagram. F: ferromagnetic phase, CAF: collinear antiferromagnetic phase,
CH: collinear helicoidal phase, H: helicoidal phase. A vertical axis at $J_2=|J_1|$ is drawn for comparison with
Fig.\ \ref{fig:quantum}.
}
\label{fig:classical}
\end{figure}

The classical $S \to \infty$ counterpart of these interactions is described by the scalar product of
commuting vectors, where the lowest energy configuration is elementarily obtained.
At any value of the exchange constants, it is described by a planar arrangement of vectors
rotated by relative angles $\vartheta_x$ in $x$ direction and $\vartheta_y$ in $y$ direction,
giving rise to the classical phase diagram in Fig.\ \ref{fig:classical} composed of four different
ordered phases:\cite{Chubukov, Rastelli}
\begin{itemize}

\item{F:} a ferromagnetic phase, $(\vartheta_x, \vartheta_y)$=$(0,0)$.

\item{CAF:} a collinear antiferromagnetic phase showing antiferromagnetic order in one direction of
the lattice and ferromagnetic order in the other one, $(\vartheta_x, \vartheta_y)$= $(0,\pi)$ or $(\pi,0)$.

\item{CH:} a collinear helicoidal phase showing helicoidal order in one direction of
the lattice and  ferromagnetic  order in the other one, $(\vartheta_x, \vartheta_y)$= $(0,q)$ or $(q,0)$
with $\cos(q) = \frac{-J_1-2J_2}{4 J_3}$.

\item{H:} a helicoidal phase composed by helicoidal order in both directions of the lattice,
$(\vartheta_x, \vartheta_y)$=$(Q,Q)$ with $\cos(Q) = \frac{-J_1}{2(J_2 +2J_3 )}$.

\end{itemize}

Opposite to the classical limit,
the quantum case $S=1/2$ has been recently analyzed using exact diagonalization (ED) techniques
to explore the complete phase diagram  in Ref.\ [\onlinecite{J1-J2-J3-shannon}],
while more detailed features of the  $J_3=0$ line were studied by ED  in
 Ref.\ [\onlinecite{J1-J2-Shannon-1, J1-J2-Shannon-2, ED-Richter}] (and also by coupled cluster methods in  Ref.\ [\onlinecite{ED-Richter}]). 
However, state of the art ED computations
can reach system sizes only up to 40 sites. Moreover, discrepancies pointed out in  Ref.\ [\onlinecite{ED-Richter}] with
earlier results call for investigating these systems with complementary techniques.

Motivated by the above discussion, we are interested in the effect of quantum fluctuations 
as introduced by large $S$ methods.
In a first step we have computed linear spin wave corrections to the classical order, as derived from
the Holstein-Primakov bosonic representation for spins.
While this method provides significant shifts for the classical antiferromagnetic $J_1-J_2-J_3$ 
 model,\cite{Jolicoeur} it gives very small corrections in the present ferromagnetic case (not reported here).
However, it helped us to localize two particular areas in the phase diagram that seem to be clearly
modified by the quantum fluctuations.
These are:
(i) the region of the classical phase transition between the ferromagnetic (F) and
collinear antiferromagnetic (CAF) phases located at $J_3$=0 and $J_2\approx 0.5 |J_1|$, where
ED studies show discrepancies,\cite{ED-Richter} and
(ii) the region $J_2\approx |J_1|$, $J_3 > 0$, where appearence of a gapped phase was reported.\cite{J1-J2-J3-shannon}

\begin{figure}[h!t]
\centering
\includegraphics[width=1\columnwidth]{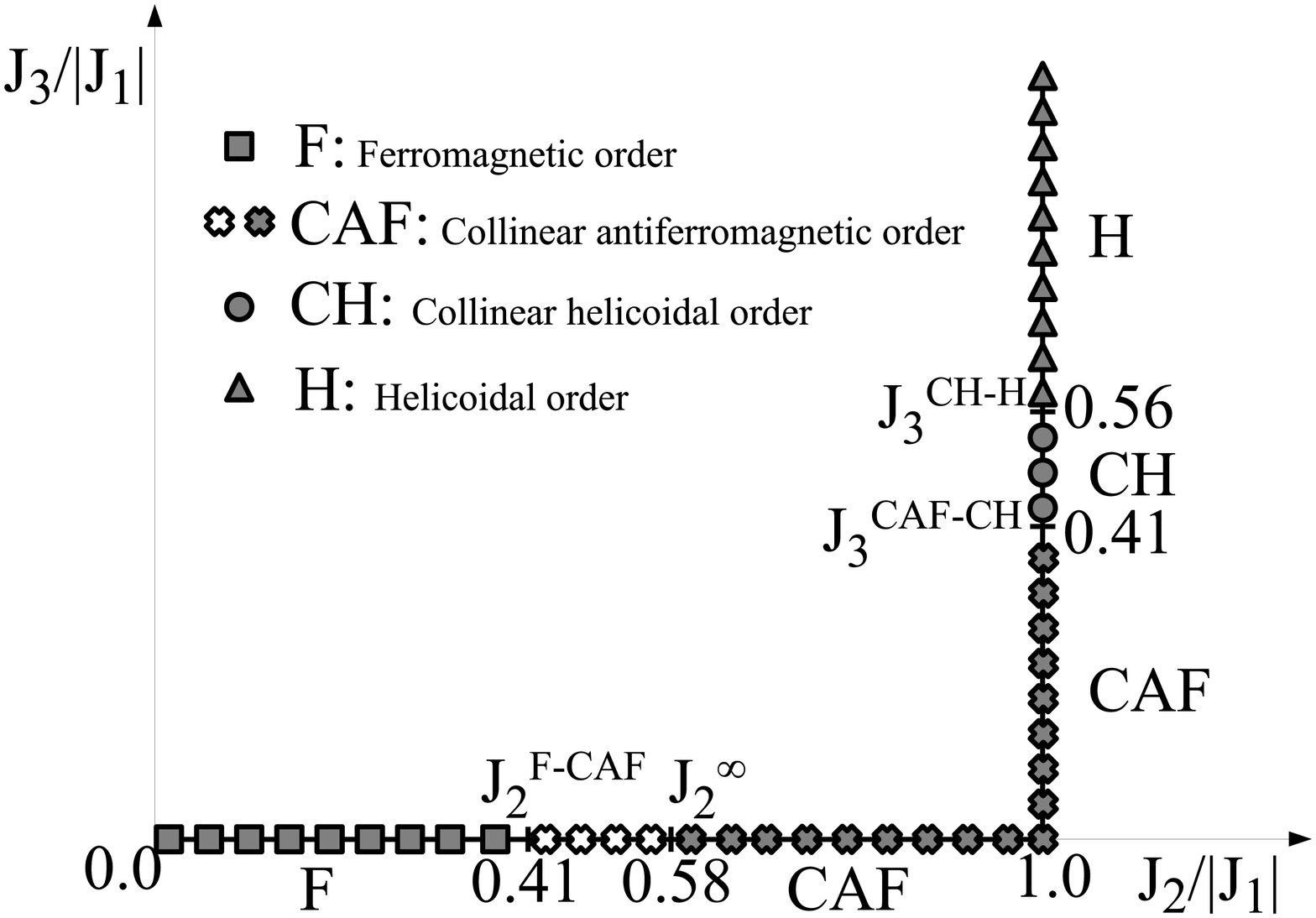}
\caption{
Corrections to the classical phase diagram computed from Schwinger boson mean field fluctuations. 
On the line $J_3=0$ we find the CAF phase for $J_2 > J_2^{F-CAF}=0.41|J_1|$; 
for $J_2 > J_2^{\infty}=0.58|J_1|$ self consistent solutions are obtained for systems up to $10^4$ sites,
while smaller system sizes are reached as $J_2$ approaches $J_2^{F-CAF}$ (empty marks). 
On the line $J_2=|J_1|$  phase boundaries are strongly shifted to larger values with respect to the classical case.
}
\label{fig:quantum}
\end{figure}

In the present work we analyze these two regions by means of the Schwinger boson (SB) representation for spins.
This method does not start from any classical order
(in contrast with linear spin wave theory) and, treated at mean field level (SBMFT), allows us to study
quantum fluctuations in fairly large systems. Moreover, it has been tested to give quite accurate results 
even for $S=1/2$, by comparison with ED (see {\it e.g.} Refs.\ [\onlinecite{ED-MFT-BS}, \onlinecite{CD}]). 
Within SB theory, long range order is characterized by boson condensation. 
We classify the different ordered phases according to the condensate momentum and the evaluation of spin correlation functions, 
finding important shifts in the classical phase boundaries. 
A schematic phase diagram with our results is 
shown in Fig.\ \ref{fig:quantum}.

The paper is organized as follows: in section \ref{sect:SBMFT} we present the SBMFT methods, in section
\ref{sect:J2} we analyze the model without $J_3$ interactions, and in section \ref{sect:J3} we analyze
the $J_2=|J_1|$ line with $J_3$ interactions. Section \ref{sect:conclusions} is devoted to the conclusions.


\section{Schwinger boson mean field theory}
\label{sect:SBMFT}

The Schwinger boson approach allows to incorporate quantum fluctuations while keeping the rotational
invariance of the Heisenberg model (see for instance Ref.\ [\onlinecite{Auerbach}]).
In this method the spin operators are written in terms of two species of bosons $b_\ua$ and $b_\da$
via the relations
\beqa
\label{eq:SB}
S_i^x=\frac{1}{2}(b^{\dag}_{i,\da}b_{i,\ua}+b^{\dag}_{i,\ua}b_{i,\da}),\nn\\
S_i^y=\frac{i}{2}(b^{\dag}_{i,\da}b_{i,\ua}-b^{\dag}_{i,\ua}b_{i,\da}),\nn\\
S_i^z=\frac{1}{2}(b^{\dag}_{i,\ua}b_{i,\ua}-b^{\dag}_{i,\da}b_{i,\da}).
\eneqa
In order to represent spin $S$ properly, one must locally fix the bosonic occupation
to 2$S$+1 states by the constraints
\beq
\label{eq:filling}
b^{\dag}_{i,\ua}b_{i,\ua}+b^{\dag}_{i,\da}b_{i,\da}=2S
\eneq
at each site $i$.

The Heisenberg Hamiltonian is then a quartic form in bosons, but can be conveniently written as
quadratic in {\it bond operators},
namely quadratic bosonic operators including one boson from each of the interacting bond sites.
Such a factorization is not unique, and different schemes are adopted in case of ferromagnetic\cite{Arovas}
or antiferromagnetic frustrated\cite{Sachdev} interactions.
A mixed scheme\cite{ED-MFT-BS, Coleman} has been shown better adapted to include both antiferromagnetic and ferromagnetic
short range correlations. It deals with more mean field parameters, but provides quantitatively better results and is our choice to study the present ferromagnetic frustrated system. 

Bond operators $A$ and $B$ are defined as
\beqa
\label{eq:AB}
&&A_{i,j}=\frac{1}{2}(b_{i,\ua}b_{j,\da}-b_{i,\da}b_{j,\ua}),\nn\\
&&B_{i,j}=\frac{1}{2}(b^{\dag}_{i,\ua}b_{j,\ua}+b^{\dag}_{i,\da}b_{j,\da}).
\eneqa
Notice that
\beqa
\label{eq:SSAB}
A^{\dag}_{i,j}A_{i,j}=\frac{1}{4}(\vec{S}_i-\vec{S}_j )^2-\frac{S}{2},\nn\\
:B^{\dag}_{i,j}B_{i,j}:=\frac{1}{4}(\vec{S}_i+\vec{S}_j )^2-\frac{S}{2},
\eneqa
where $:\mathcal{O}:$ means the bosonic normal order of an operator $\mathcal{O}$, relate non-vanishing $A$ to antiferromagnetic structures and  non-vanishing $B$ to ferromagnetic structures.
Moreover, expanding the squares yields representations for the $SU(2)$ invariant terms $\vec{S}_i \cdot \vec{S}_j$.
The Hamiltonian (\ref{eq:H}) can then be written as
\beq
\label{eq:HAB}
H=\sum_{r=1,2,3} J_r \sum_{\langle i,j \rangle_r} (:B^{\dag}_{i,j}B_{i,j}:-A^{\dag}_{i,j}A_{i,j})
+H_{\lambda}
\eneq
where the term
\beq
\label{eq:constraint}
H_{\lambda}=\sum_i \lambda_i (b^{\dag}_{i,\ua}b_{i,\ua}+b^{\dag}_{i,\da}b_{i,\da}-2S)
\eneq
forces the local constraints, $\lambda_i$ being the Lagrange multipliers.

At mean field level, we perform a Hartree-Fock decoupling introducing a uniform Lagrange multiplier $\lambda$
and translationally invariant parameters $\alpha, \beta$ for the expectation values of each type of bond operator present in the Hamiltonian 
(as mentioned above, this decoupling is not unique). 
As it is known,\cite{ED-MFT-BS,Arovas,Sachdev,Coleman} the most severe approximation here is the violation of the local boson number constraint
in Eq.\ (\ref{eq:filling}), which is only respected on average.
We are thus dealing with the Lagrange multiplier $\lambda$, six $\alpha$'s and six $\beta$'s as independent variational parameters, 
the latter set as expectation values of bond operators
\beq
\begin{array}{ll}
\alpha_1=\langle A_{\vec{r},\vec{r}+\breve{x}} \rangle,  &   \tilde\alpha_1=\langle A_{\vec{r},\vec{r}+\breve{y}}\rangle,  \\
\alpha_2=\langle A_{\vec{r},\vec{r}+\breve{x}+\breve{y}} \rangle,  &  \tilde\alpha_2=\langle A_{\vec{r},\vec{r}+\breve{x}-\breve{y}}\rangle, \\
\alpha_3=\langle A_{\vec{r},\vec{r}+2\breve{x}} \rangle,  &  \tilde\alpha_3=\langle A_{\vec{r},\vec{r}+2\breve{y}} \rangle, 
\end{array}
\label{eq:mfp}
\eneq
and similar expresions relating the $\beta$'s to $\langle B_{i,j}\rangle$ expectation values.
For compact notation we write  $\alpha_{i,j}=\langle A_{i,j}\rangle$,
$\beta_{i,j}=\langle B_{i,j}\rangle$, using site indices to indicate the range of the bond $\langle i,j \rangle_r$ ($r=1,2,3$)
as well as the possible orientations along the lattice described in Eq.\ (\ref{eq:mfp}). 
The mean field Hamiltonian then reads
\beqa
H_{MF} & = & \sum_{r=1,2,3} J_r \sum_{\langle i,j \rangle_r}
    \left( \beta^*_{i,j} B_{i,j} + B^{\dag}_{i,j}\ \beta_{i,j}+\right. \nn\\
& &   \left. -\alpha^*_{i,j} A_{i,j}- A^{\dag}_{i,j} \alpha_{i,j} \right) \nn\\
& & -\sum_{r=1,2,3} J_r \sum_{\langle i,j \rangle_r}
(|\beta_{i,j}|^2-|\alpha_{i,j}|^ 2)
+ H_{\lambda}.
 \label{eq:HBS-MF}
\eneqa
After a Fourier transform one gets momentum modes with quadratic  terms which are not particle
number conserving. These are diagonalized by a standard Bogoliubov transformation, depending on the
variational parameters and rendering decoupled modes
with simple particle number conserving, positive, quadratic, Hamiltonian
%
%
\beq
H_{MF}= \sum_{\vec{k}}\left[\omega(\vec{k}) \eta^\dag_{\vec{k}} \eta_{\vec{k}}
\right]+const,
\label{eq:HBog}
\eneq
where
\beq
\eta_{\vec{k}}=
\left(\begin{array}{c} 
d_{\vec{k}\ua}  \\
d_{-\vec{k}\da}^\dag  
\end{array}\right)
\label{eq:dBog}
\eneq
contains the Bogoliubov bosonic operators, with dispersion relation $\omega(\vec{k})$, and $const$ stands for non operator terms.

Finally, we compute self consistently the mean field parameters by minimizing the ground state (Bogoliubov vacuum) energy
with respect to $\lambda$ and equating $\alpha_{i,j}$ and $\beta_{i,j}$ with the ground state expectation values of the
corresponding operators.
Such computation is done numerically on finite lattices of $N$ sites with periodic boundary conditions,
allowing the study of large system sizes, up to $10^4$ sites in the present work.

We must stress that our procedure is not suited for the ferromagnetic phase, where parameters $\alpha_{i,j}$
vanish and the Hamiltonian in Eq.\ (\ref{eq:HBS-MF}) is already particle conserving: the Schwinger
bosons vacuum simply violates the constraint in Eq.\ (\ref{eq:filling}), even on average.
We use then the exact energy of a fully polarized (ferromagnetic) state, $E_{F} = 2 N S^2 (J_1+J_2+J_3)$, 
for comparison with SBMFT energies, or extrapolations thereof, to determine the ferromagnetic phase boundaries.

Once the self consistent equations are solved, the tools above allow to compute any kind of
observable on the ground state.
In the present work we have set $S=1/2$ and studied four quantities: the dispersion relation, its gap, the modulated
magnetization $M_n^2$ (defined below) and the spin correlation function.

\begin{figure}[h!t]
\centering
\includegraphics[width=1\columnwidth]{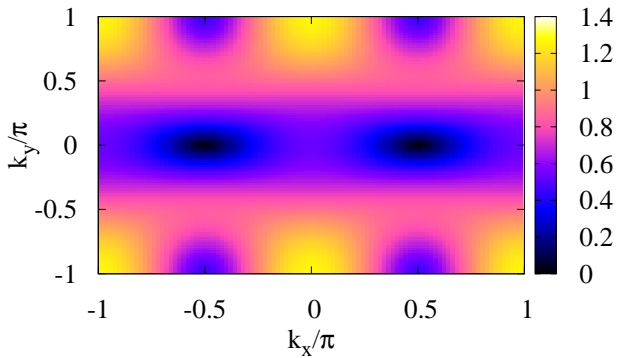}
\caption{(color online) Dispersion relation for $J_2=0.75|J_1|$ and $J_3$=0 and for a system of size $N=100\times$100
(in arbitrary scale, with darker zones indicating lower energy).
The bosons condense at  points $k_x$=$\pm \pi/2$, $k_y$=0,  which correspond to a CAF phase $(\pm \pi,0)$.}
\label{fig:dispersion}
\end{figure}

When the dispersion relation shows a zero mode, Bose condensation indicates an ordered phase,
in the sense that the spin structure factor shows a maximum at a pitch angle
$\vec{\theta}=(\theta_x,\theta_y)$ commensurate with the finite lattice,
related to the position of the zero mode of the dispersion relation $\vec{k}_{min}$
by $\vec{\theta}=2\vec{k}_{min}$.\cite{k-q}
Notice that Bose condensation depends on boson density, related in the SB approach to the spin $S$
representation by the constraint in Eq.\ (\ref{eq:filling}).
As we study numerically the lowest density case, $S=1/2$, such ordered phases will also be present for larger $S$.
For illustration purpose, in Fig.\ \ref{fig:dispersion} we show the dispersion relation at coupling
values $J_2=0.75|J_1|$ and $J_3=0$ (well inside the CAF classical phase).

In the case of long range order, the vanishing of the gap in
the thermodynamical limit is usually recovered after a finite size scaling analysis.
Another issue arising at finite sizes is that related to commensurability.
As the gap is obtained through the value of the minima of the dispersion relation on
the reciprocal lattice, in the case that a thermodynamical minimum does not 
fit with the available momenta values at finite sizes, 
numerical difficulties may show up (see section \ref{sect:J3}).

Regarding the other observables, the spin correlations $\langle\vec{S}_i\cdot\vec{S}_j\rangle$
for any pair of sites $i,j$ are computed on the ground state from the bosonic representation
in Eq.\ (\ref{eq:SB}) and the Bogoliubov transformation obtained numerically. 
%
We measure modulated magnetizations by considering the $\vec{\theta}$-dependent susceptibility $M_n^2(\vec{\theta})$,
defined as the following average over the lattice:
\beq
M_n^2(\vec{\theta})=\frac{1}{N(N+2)}\sum_{i,j}\langle\vec{S_i}.\vec{S_j}\rangle e^{i\vec{\theta}.(\vec{R_i}-\vec{R_j})},
\eneq
where $\vec{\theta}=(\theta_x,\theta_y)$ and $\vec{R_i}$ is the position of spin $\vec{S_i}$ with respect
to some reference site. It amounts to an improveded\cite{Schulz} normalization of the spin structure factor,
that fits better small systems and tends to moderate the weight of strong on-site terms.
It is straightforwardly computed from the spin correlations.


\section{Nearest neighbors frustration $J_2$}
\label{sect:J2}

In this section we analyze the case $J_3=0$, that is a system with ferromagnetic first neighbors
couplings and  only second neighbors antiferromagnetic interactions.

The classical phases on this line, shown in Fig.\ \ref{fig:classical}, are F and CAF, separated by a critical
value ${J_2}^{class}=0.5 |J_1|$.
The quantum case was studied for $S=1/2$: based on ED of the model and coupled cluster methods, 
Richter {\it et. al} \cite{ED-Richter} predict a simple shift of the critical coupling to {\it lower}
$J_2 = 0.39 |J_1|$,
while Shannon  {\it et al.} \cite{J1-J2-Shannon-1, J1-J2-Shannon-2} estimate by ED a CAF phase only for
{\it larger} $ J_2 \gtrsim 0.6 |J_1|$ and predict the presence of a
quadrupolar (bond-nematic) phase in the critical area, $0.4 \lesssim J_2/|J_1| \lesssim 0.6$.

We have studied with SBMFT systems of sizes ranging from  $N=4\times 8$ (finding excellent agreement with ED, for ground state energies) 
up to $N$=100$\times$100.

\begin{figure}[h!t]
    \centering
        \includegraphics[width=1\columnwidth]{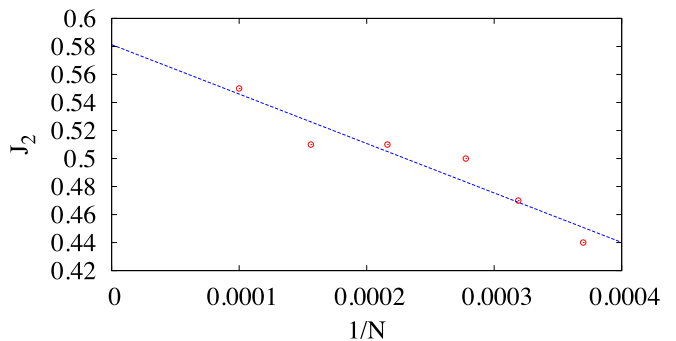}
	\caption{
	Evolution of the lowest couplings $J_2(N)$ tractable within SBMFT along the line $J_3$=0 with the inverse of the system size.
	Temptative linear extrapolation realized for sizes $N\times N$ with $N=52, 56, 60, 68, 80, 100$. 
	}
    \label{fig:J2cCAF}
\end{figure}

As mentioned in Sect.\ \ref{sect:SBMFT}, our procedure does not provide a self-consistent solution for the
F phase. 
We first analyze the values  of $J_2$ above which SBMFT solutions are obtained.
These values turn out to be sensitively dependent on the system size. 
Above $J_2 \approx 0.56|J_1|$ we reach solutions for all explored systems, up to $100\times 100$ sites; 
but approaching the F phase we get oscillatory behaviour with the system trapped in metastable configurations, and the tractable sizes reduce as down as $20\times 20$ at $J_2 \approx 0.4|J_1|$.
The size dependence of the lowest couplings $J_2(N)$ tractable within SBMFT  is roughly linear in $1/N$, as shown in
Fig.\ \ref{fig:J2cCAF}, suggesting an infinite size extrapolation to ${J_2}^{\infty}=0.58\,|J_1|$. 
Thus we estimate that, investing enough CPU time, one can treat systems of arbitrary size only when $J_2>{J_2}^{\infty}$.
%

For $J_2$ above $J_2^{\infty}$
the observables computed from the SBMFT self-consistent solutions
correspond to a CAF phase, showing staggered magnetization along one of the lattice axes.
An example of the dispersion relation,
for $J_2=0.75\,|J_1|$ and $J_3$=0 in a large system of size $N=100\times100$,
is the one shown in Fig.\ \ref{fig:dispersion}.
The boson modes become gapless at momentum points $\vec{k}=(\pm\pi/2,0)$,
showing that the bosons do condense.
The condensation momenta correspond to ordering angles $\vec{\theta}=(\pm\pi,0)$.
The same pattern (alternatively with $\vec{\theta}=(0, \pm\pi)$) is found for $J_2 > J_2(N)$, $N = 20, 40, 60, 80, 100$.

\begin{figure}[h!t]
    \centering
        \includegraphics[width=1\columnwidth]{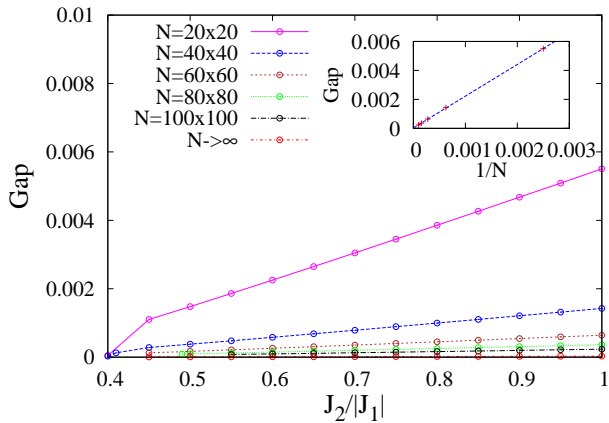}
	\caption{(color online) Energy gaps (minima of the dispersion relation) for systems of different size,
	in the CAF phase. Extrapolated value of the gap vanishes, confirming collinear antiferromagnetic order. 
  Inset: gap extrapolation for $J_2=|J_1|$.
	}
    \label{fig:CAF-gaps}
\end{figure}

\begin{figure}[h!t]
    \centering
        \includegraphics[width=1\columnwidth]{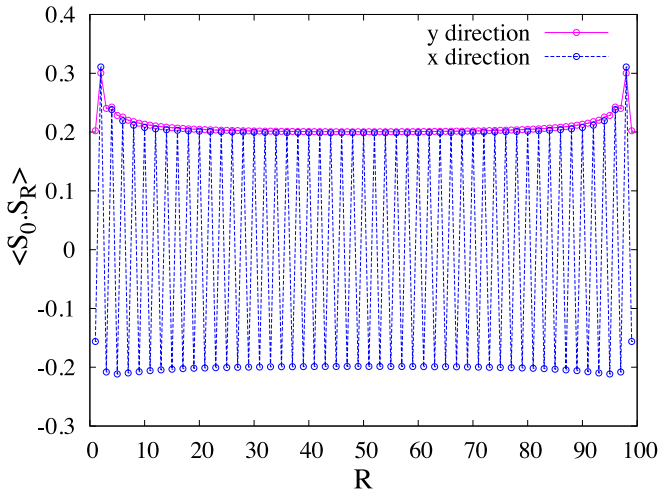}
	\caption{(color online) Spin correlation function for a system of size $N=100\times100$ at
	$J_2=0.75$, $|J_1|$ and $J_3$=0 (corresponding to the dispersion in Fig.\ \ref{fig:dispersion}), along both lattice axes.
	}
    \label{fig:CAF-correlations}
\end{figure}

\begin{figure}[h!t]
    \centering
        \includegraphics[width=1\columnwidth]{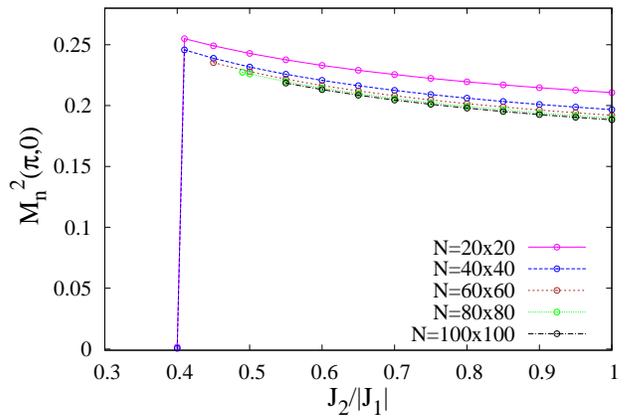}
	\caption{(color online) Staggered magnetization as order parameter for the CAF phase, at $J_3=0$.
	}
    \label{fig:CAF-Mn2}
\end{figure}

The dispersion relation gap goes clearly to zero for $N\rightarrow\infty$, as shown in Fig.\ \ref{fig:CAF-gaps} 
(size scaling for $J_2=|J_1|$ is shown in the inset).
Correspondingly, the spin correlation function exhibits long range order:
when  $\vec{\theta}=(\pm\pi,0)$, we observe antiferromagnetic correlations in the $x$ direction
and ferromagnetic correlations in the  $y$ direction. For example,
in Fig.\ \ref{fig:CAF-correlations} we show the correlations for $J_2=0.75|J_1|$.
The corresponding modulated magnetization has a maximum at $(\pm\pi,0)$.
Then $M_n^2(\pi,0)$,  shown in Fig.\ \ref{fig:CAF-Mn2}, measures the staggered magnetization along the $x$-direction. 
In general, $M_n^2(\pi,0)$ and $M_n^2(0,\pi)$ can be used as order parameters for the CAF phase.

We now turn to discuss the region $0.4 \lesssim J_2/|J_1| \lesssim 0.58$, under controversy in the literature, where the largest sizes studied do not provide a self-consistent solution.
Notice in Figs.\ \ref{fig:CAF-gaps} and \ref{fig:CAF-Mn2} that in all converged solutions
there is no signal of an exotic phase but clear indications of the CAF phase.
It turns out that the finite size available solutions scale in this range to the same CAF phase as in the range  ${J_2}^{\infty}<J_2<|J_1|$.

\begin{figure}[h!t]
    \centering
        \includegraphics[width=1\columnwidth]{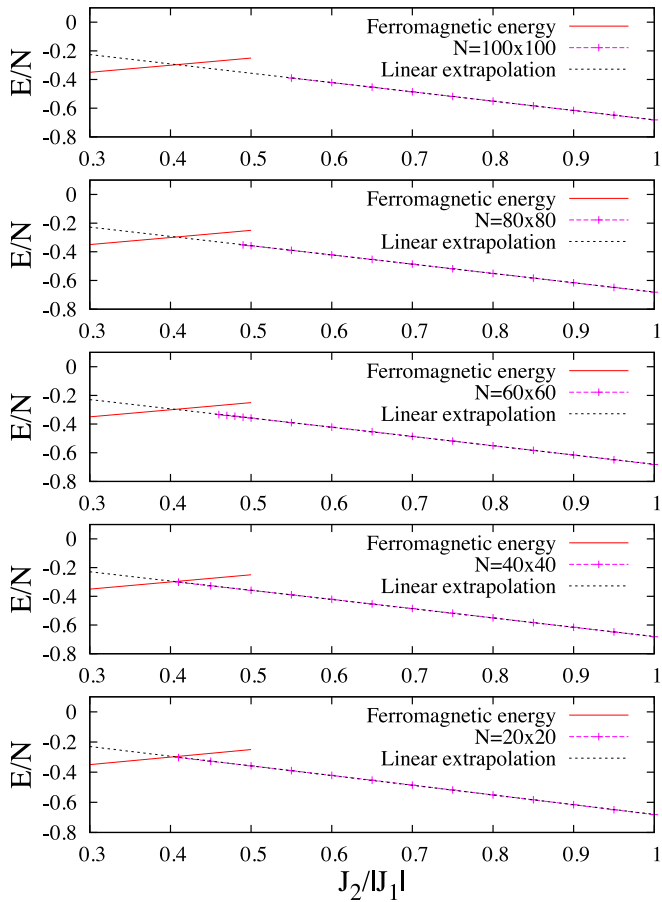}
	\caption{(color online) Ground state energy in the CAF phase, for the same system sizes shown in Fig.\ \ref{fig:J2cCAF}.
	Smaller sizes (lower panels) show energies obtained for couplings down to $J_2/|J_1|\sim 0.4$, in full agreement with  extrapolated energies from the region  $J_2/|J_1| >0.58$.
	Exact ferromagnetic phase energy (solid line) is included to illustrate the proposed F-CAF phase boundary.
	}
    \label{fig:CAF-Egs}
\end{figure}

\begin{figure}[h!t]
    \centering
        \includegraphics[width=1\columnwidth]{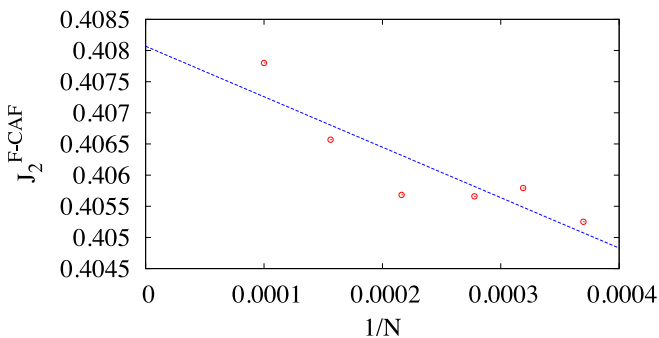}
	\caption{Size dependence of  ${J_2}^{F-CAF}(N)$, the crossing point of ferromagnetic phase energy
	and the extrapolated CAF energy (smallest sizes not shown here).
	}
    \label{fig:F-CAF-crossing}
\end{figure}

The energy per site obtained from SBMFT self-consistent solutions is very stable against
system size and shows a neat linear dependence with $J_2$, including the region $0.4 \lesssim J_2/|J_1| \lesssim 0.58$.
(see Fig.\ \ref{fig:CAF-Egs}). 
We conclude that lack of convergence of SBMFT self-consistent equations at large system size
is an artifact of our present approach, presumably due to the proximity of a ferromagnetic phase, 
and that solutions obtained for small systems are good estimates of the CAF phase in the region for
$0.4 \lesssim J_2/|J_1| < 0.58$.
Then, following the criteria in Ref.\ [\onlinecite{ED-Richter}] we set
the F-CAF phase boundary at the intersection point between the extrapolated CAF energy and
the exact energy per site of a fully polarized state, $E_{F}/N = 2 S^2 (J_1+J_2)$ (here with $S=1/2$).
Such a point ${J_2}^{F-CAF}(N)$ appears, depending on size $N$, at a quite precise value of $J_2/|J_1|$ between
$0.402$ and $0.4078$, as shown in Fig.\ \ref{fig:F-CAF-crossing}.
The roughly linear dependence in $1/N$ suggests an extrapolated transition point at ${J_2}^{F-CAF}=0.41\,|J_1|$.

From the results in this Section we conclude that, on the $J_3=0$ line,
the present SBMFT approach can confirm the CAF phase for $J_2 > {J_2}^{F-CAF}=0.41\,|J_1|$.
For $J_2 > {J_2}^{\infty}=0.58\,|J_1|$, this can be tested even in the thermodynamical limit.
For lower $J_2$, although systems of limited size can be solved, we find CAF observables until 
the transition to the F phase.
A linear extrapolation of the CAF ground state energies suggests a direct first order transition to the F phase at
$J_2^{F-CAF}=0.41\,|J_1|$ in accordance with Ref.\ [\onlinecite{ED-Richter}].


\section{Effects of next to nearest neighbor frustration $J_3$}
\label{sect:J3}

In this section we analyze the influence of third neighbors antiferromagnetic
couplings $J_{3}$ on top of the CAF phase, by fixing $J_{2}=|J_{1}|$.
We recall that, on this line, the classical phase diagram in Fig.\ \ref{fig:classical} shows collinear
antiferromagnetic order for $0<J_{3}<0.25|J_{1}|$, a continuous transition
to collinear helicoidal $(q,0)$ (or $(0,q)$) order for $0.25|J_{1}|<J_{3}<0.5|J_{1}|$,
with $q$ decreasing from $\pi$ to $\frac{2}{3}\pi$, and a discontinuous
transition to a helicoidal phase $(Q,Q)$ for $J_{3}>0.5|J_{1}| $, with $Q$
increasing in a narrow window, from $0.4195\pi$ to $\frac{\pi}{2}$
(reaching $Q=0.4466\pi$ at $J_{3}=|J_{1}|$, the largest value of $J_{3}$ in the present analysis). 

The quantum case was studied by Sindzingre {\it et al.} 
\cite{J1-J2-J3-shannon} by ED in systems up to $N=36$ sites, for
positive $J_{2}$ and $J_{3}$, both up to $|J_{1}|$. 
For $J_{2}> 0.75|J_{1}|$ and around the classical boundary between collinear helicoidal and helicoidal phases,
the authors find signals of an exotic gapped phase, stating that is difficult to conclude its precise nature because of large and irregular finite size effects. In particular, on the line
$J_{2}=|J_{1}|$, they find a CAF phase
for $J_{3} \lesssim 0.35|J_{1}|$ and a gapped phase for $J_{3} > 0.35|J_{1}|$.

We have applied the SBMFT to systems of size $N=20\times20,\,40\times40,\,60\times60,\,80\times80,\,100\times100$.
For $0<J_{3} \lesssim 0.4|J_{1}|$, we find persistence of the $CAF$
phase: the dispersion relation remains gapless at commensurate momenta
$\vec{k}=(\pm\pi/2,0)$ (or $(0,\pm\pi/2)$) and the $\vec{\theta}$-dependent susceptibility has
a maximum at $(\pm\pi,0)$ (or $(0,\pm\pi)$). This phase shows a boundary 
that barely depends on system sizes and can be estimated as ${J_3}^{CAF-CH} \approx 0.41|J_{1}|$. 
This amounts to a shift of $0.16|J_1|$ with respect to the classical value.

\begin{figure}[h!t]
\centering
\includegraphics[width=1\columnwidth]{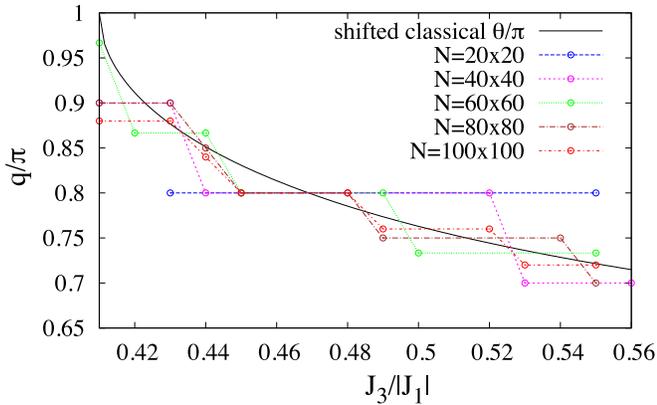}
\caption{(color online) Evolution of the incommensurate $q$ along the line $J_2$=$|J_1|$, in the collinear helicoidal phase
for different system sizes. 
For comparison, we show the classical pitch angle $\vartheta(J_3-0.16|J_1|)$ ({\em i.e.} plotted 
with respect to the SBMFT phase boundary ${J_3}^{CAF-CH}$). 
} 
\label{fig:q_evolution}
\end{figure}

For larger $J_{3}$ the minima of the dispersion relation move to incommensurate
values of $\vec{k}$. It gets numerically difficult on a finite
lattice to determine the existence of gapless minima. However, though
with less precision than in the CAF phase, we find for all studied
sizes that, immediately above ${J_3}^{CAF-CH}(N)$, a gapless collinear
helicoidal phase with $\vec{k}=(q,0)$ (or $(0,q)$) develops. The available values
for $q$ are discrete but, as shown in Fig. \ref{fig:q_evolution}, the dispersion minima
position evolves in the same range as the classical ones, simply
shifted in $J_{3} \to J_{3}+0.16|J_1|$. For each point in the figure, the $\vec{\theta}$-dependent susceptibility
shows a maximum at $\vec{\theta}=(q,0)$ (or $(0,q)$), characterizing collinear helicoidal magnetization
order.
%

The collinear helicoidal phase extends up to ${J_3}^{CH-H} \approx 0.56 |J_1|$. Again,
such boundary is almost independent of the system sizes. 
For even larger $J_{3}$ further numerical difficulties show up. Indeed, the dispersion seems to get
gapless at momenta $(Q,Q)$ with $Q$ in the same narrow window found in the classical phase diagram for 
the helicoidal phase. As the thermodynamical values of minima position in such a narrow range may mismatch the available momenta 
for a given finite lattice, it is difficult to select the minimum of the dispersion relation amongst neighboring points. 
For this same reason, a gap may seem to open but could be just a finite size artifact.
A refinement was done by choosing different system sizes in order to allow for different distributions of lattice momenta.
In general, when a clear minimum is found, it corresponds to a gapless situation with ordering angle 
$\vec{\theta}=(Q,Q)$ where $0.421\pi < Q < 0.483\pi$. In such cases, the order parameter $M_n^2(Q,Q)$ 
remains finite signaling a helicoidal phase. Thus, the present method shows no clear indications of the 
appearance of a gap in the excitation spectrum.


\section{Conclusions}
\label{sect:conclusions}

To summarize, along the line $J_3$=0 and within SBMFT we can confirm the collinear antiferromagnetic phase for 
$J_2 > {J_2}^{F-CAF}=0.41\,|J_1|$. For ${J_2}^{F-CAF} < J_2 < {J_2}^{\infty}=0.58|J_1|$ the convergence becomes harder,
presumably because of dominance of ferromagnetic correlations. 
However, at tractable system sizes we find CAF observables until the transition to the ferromagnetic phase.
A linear extrapolation of the CAF phase ground state energies
suggests a direct first order transition to the ferromagnetic phase at
${J_2}^{F-CAF}$ in good agreement with Ref.\ [\onlinecite{ED-Richter}].

Along the line $J_2$=$|J_1|$ we have found that the boundaries between the collinear and 
incommensurate phases are strongly shifted, with respect to the classical case, to larger values of $J_3$:
${J_3}^{CAF-CH} \approx 0.41|J_{1}|$
and ${J_3}^{CH-H} \approx 0.56 |J_1|$.
We do not find clear evidence of spin-gapped phases within the present approximation.  

\begin{acknowledgments}
We thank A.\ Honecker, C.A.\ Lamas and H.D.\ Rosales for fruitful discussions.
H.F.\ thanks DAAD for grant A/10/70636.
D.C.\ and G.R.\ are partially supported by CONICET (PIP 1691) and ANPCyT (PICT 1426).
\end{acknowledgments}




\end{document}